\documentclass[aps,prd,amsmath,amssymb,preprintnumbers,preprint,nofootinbib,a4paper,11pt]{revtex4-1}
\pdfoutput=1
\usepackage{graphicx}
\graphicspath{{./figures/}}
\usepackage{url}
\usepackage[bookmarks, pagebackref=true]{hyperref}
\usepackage{color}
	\definecolor{rossoCP3}{cmyk}{0,.88,.77,.40}
		\definecolor{graa}{rgb}{0.8,0.8,0.8}
		\definecolor{blaa}{rgb}{0.2,0.2,0.6}
		\hypersetup{
			colorlinks, 
			bookmarksopen, 
			bookmarksnumbered,
			citecolor=blaa, 		
			linkcolor=rossoCP3,	
			urlcolor=rossoCP3,			
			}
\usepackage{bm}
\usepackage{bbm}
\usepackage{pxfonts}
\usepackage[margin=5pt, font=small,labelfont=bf,justification=raggedright]{caption}
\usepackage{subfig}
\usepackage{youngtab}
\usepackage{slashed}
\usepackage{bbold}

\newcommand{\beq}{\begin{eqnarray}}
\newcommand{\eeq}{\end{eqnarray}}

\newcommand{\ea}[1]{
\begin{align}
#1
\end{align}
}

\newcommand{\bmp}{\noindent\begin{minipage}{16cm}}
\newcommand{\emp}{\end{minipage}\vskip 7mm} 

\begin{document}
\phantom{g}\vspace{2mm}
\title{\texorpdfstring{ \LARGE  \color{rossoCP3} 
Standard Model-like corrections to Dilatonic Dynamics}{Standard Model-like corrections to Dilatonic Dynamics}}
\author{Oleg {\sc Antipin} 
}\email{antipin@cp3-origins.net}
\author{Jens {\sc Krog}
}\email{krog@cp3-origins.net} 
\author{Esben {\sc M\o lgaard}
}\email{molgaard@cp3-origins.net} 
\author{Francesco {\sc Sannino}
}\email{sannino@cp3.dias.sdu.dk} 
\affiliation{
{ \color{rossoCP3}  \rm CP}$^{\color{rossoCP3} \bf 3}${\color{rossoCP3}\rm-Origins} \& the Danish Institute for Advanced Study {\color{rossoCP3} \rm DIAS},\ \\ 
University of Southern Denmark, Campusvej 55, DK-5230 Odense M, Denmark.
}
\begin{abstract}
We examine the effects of standard model-like interactions on the near-conformal dynamics of a theory featuring a dilatonic state identified with the standard model-like Higgs. As template for near-conformal dynamics we use a gauge theory with fermionic matter and elementary mesons possessing the same non-abelian global symmetries as a technicolor-like theory with matter in a complex representation of the gauge group. We then embed the electroweak gauge group within the global flavor structure and add also ordinary quark-like states to mimic the effects of the top. We find that the standard model-like induced corrections modify the original phase diagram and the details of the dilatonic spectrum. In particular, we show that the corrected theory exhibits near-conformal behavior for a smaller range of flavors and colors. For this range of values, however, our results suggest that near conformal dynamics could accommodate the observed  Higgs-like properties. \\[.1cm]
{\footnotesize  \it Preprint: CP$^3$-Origins-2013-6 \& DIAS-2013-6}

\end{abstract}
\maketitle

\newpage

\section{Introduction}

The lightness of the Higgs-like state discovered at the Large Hadron Collider (LHC) experiments \cite{Chatrchyan:2012ufa,Aad:2012tfa} with respect to the natural scale of the standard model which is $4\pi v$ with $v\simeq 246$~GeV has led to speculation that the Higgs-like state is a protected state, perhaps associated to the spontaneous breaking of a dilatonic symmetry. 
 
The prototype of this kind of models would require the construction of a perturbative \cite{Grinstein:2011dq, Antipin:2011aa,Lalak:2012ax} or nonperturbative \cite{Holdom:1981rm,Sannino:1999qe,Hong:2004td,Dietrich:2005jn,Sannino:2009za,Bellazzini:2012vz,Campbell:2011iw}  quantum field theory possessing near-conformal dynamics directly coupled to the standard model fields. We refer to \cite{Jarvinen:2009fe,Antipin:2009dz,Alanen:2010tg,Arean:2012mq,Elander:2009pk,Evans:2013vca} for holographic constructions. Here we wish to analyse the corrections to the fixed point structure and dynamics of the underlying gauge dynamics stemming from the standard model sector. 

To be quantitative we use as prototype of near conformal gauge dynamics one in which the infrared stable (IR) fixed points are accessible in perturbation theory with the further property that spontaneous symmetry breaking and consequent loss of conformality occur simultaneously because of quantum corrections to the potential of the theory \cite{Antipin:2011aa}. The model posseses the same non-abelian global symmetry as a technicolor like theory with $N_f$ flavors. It is therefore natural to gauge $N_f/2$ doublets with respect to the weak interactions. To mimic the effects of the standard model top quark we introduce quark-like states coupling via Yukawa interactions to the elementary mesonic sector of the theory. 
 
Taking into account the new standard model-like fields and interactions with the perturbative technicolor-like sector we re-determine the underlying model's beta functions, fixed points, phase structure, spectrum, and finally compare it to the original model.

We show that the corrections modify the original phase diagram and the details of the dilatonic spectrum. The standard model-like corrected theory still exhibits a near-conformal behavior but for a smaller range of flavors and underlying colors. However, for this range of values the results indicate that near conformal dynamics could accommodate the observed  Higgs-like properties. Attention must be paid in interpreting our results within the composite Higgs model scenarios. Here we should consider also the electroweak corrections to the composite Higgs mass squared proportional to the cutoff squared of the model tending to further reduce its physical mass \cite{Foadi:2012bb} due to the top corrections.
 
The paper is structured as follows. In section \ref{amsSec}, we briefly summarise the model featuring near conformal dynamics and a dilaton in perturbation theory \cite{Antipin:2011aa}.  Then, in section \ref{smImp} we embed the electroweak theory and introduce the standard model quark-like states.  The corrections on the original dynamics are then presented in section \ref{smCor}. Here we determine the modified phase diagram, discuss the corrected dilatonic spectrum and comment on the phenomenological impact of our results. We finally conclude  in section \ref{conclusion}.

\section{A light dilaton dynamics resembling walking technicolor}\label{amsSec}
The setup of the original model \cite{Antipin:2011aa} features a QCD-like sector with gauge coupling $g$, elementary mesons, a fermion in the adjoint representation of the $SU(N_{TC})$ gauge group and a Yukawa interaction between the scalar singlets and fermions in the fundamental representation. The Lagrangian is 
\begin{align}
\label{amsL}
\mathcal{L} = \mathcal{L}_K(G_{\mu},\lambda_m,Q,\tilde{Q},H) + \left(y_H QH\tilde{Q} + \text{h.c}\right) 
		      -u_1\left( \text{Tr}\,[HH^{\dag}] \right)^2 - u_2\text{Tr}\left[(HH^{\dag})^2 \right].
\end{align}
The symmetries of the theory are given in Table \ref{amssym}.

\begin{table}[h]
   \[
   \begin{array}{c|c|c c c c}
     \hline
     \hline
     Fields & [SU(N_{TC})] & SU(N_f)_L & SU(N_f)_R & U(1)_V & U(1)_{AF}\\ \hline 
     \lambda_m & Adj &$1$&$1$&$0$ & $1$ \\
     Q &  \Box  & \overline{\Box} & 1 & \tfrac{N_f-N_{TC}}{N_{TC}} & -\tfrac{N_{TC}}{N_f} \\ 
     \tilde{Q} & \overline{\Box}  & 1 & \Box & -\tfrac{N_f-N_{TC}}{N_{TC}} & -\tfrac{N_{TC}}{N_f} \\
     \hline
     H & 1 & \Box & \overline{\Box} & 0 & \tfrac{2N_{TC}}{N_f} \\
     G_{\mu} & Adj & 1 & 1 & 0 & 0 \\ \hline \hline
   \end{array}
   \]
  \caption{The field content of the model and the related symmetries}
  \label{amssym}
\end{table}

This model features nontrivial dynamics due to the presence of calculable fixed points of the perturbative beta functions.
These fixed points are perturbative when $N_f$ is in the region just below $\tfrac{9}{2}N_{TC}$, which will be controlled through the parameter $x\equiv N_f/N_{TC}$. In the original analysis, it was found to be useful to consider the limit where $N_f,N_{TC} \rightarrow \infty$ while $x$ is kept fixed. We will not consider this limit, but for easy comparison, we will keep the same notation. In particular, we rescale the couplings as follows
\begin{equation}
a_g = \frac{g^2N_{TC}}{(4\pi)^2},\quad a_H = \frac{y_h^2N_{TC}}{(4\pi)^2},\quad z_1 = \frac{u_1N_f^2}{(4\pi)^2},\quad  z_2 = \frac{u_2N_f}{(4\pi)^2}.
\end{equation}

With these definitions, the beta functions are found to one loop order in the quartic and Yukawa couplings and to two loop order for the gauge coupling:
\begin{align}
\beta_{a_g} &= -2a_g^2\left[3-\frac{2x}{3}+x^2a_H+ \left(\frac{x^3}{N_f^2}+6-\frac{13x}{3}\right)a_g \right] \label{amsbetag}\\
\beta_{a_H} &= 2a_H\left[(1+x)a_H-3\left(1-\frac{x^2}{N_f^2}\right)a_g \right] \label{amsbetaH}\\
\beta_{z_1} &= 4\left[\left(1+\frac{4}{N_f^2}\right)z_1^2+4z_1z_2+3z_2^2+z_1a_H\right] \label{amsbeta1}\\
\beta_{z_2} &= 4\left[\frac{6}{N_f^2}z_1z_2 + 2z_2^2+z_2a_H-\frac{x}{2}a_H^2\right]\label{amsbeta2} 
\end{align}

\begin{figure}[h]
\begin{center}
\includegraphics[width = 0.6\linewidth]{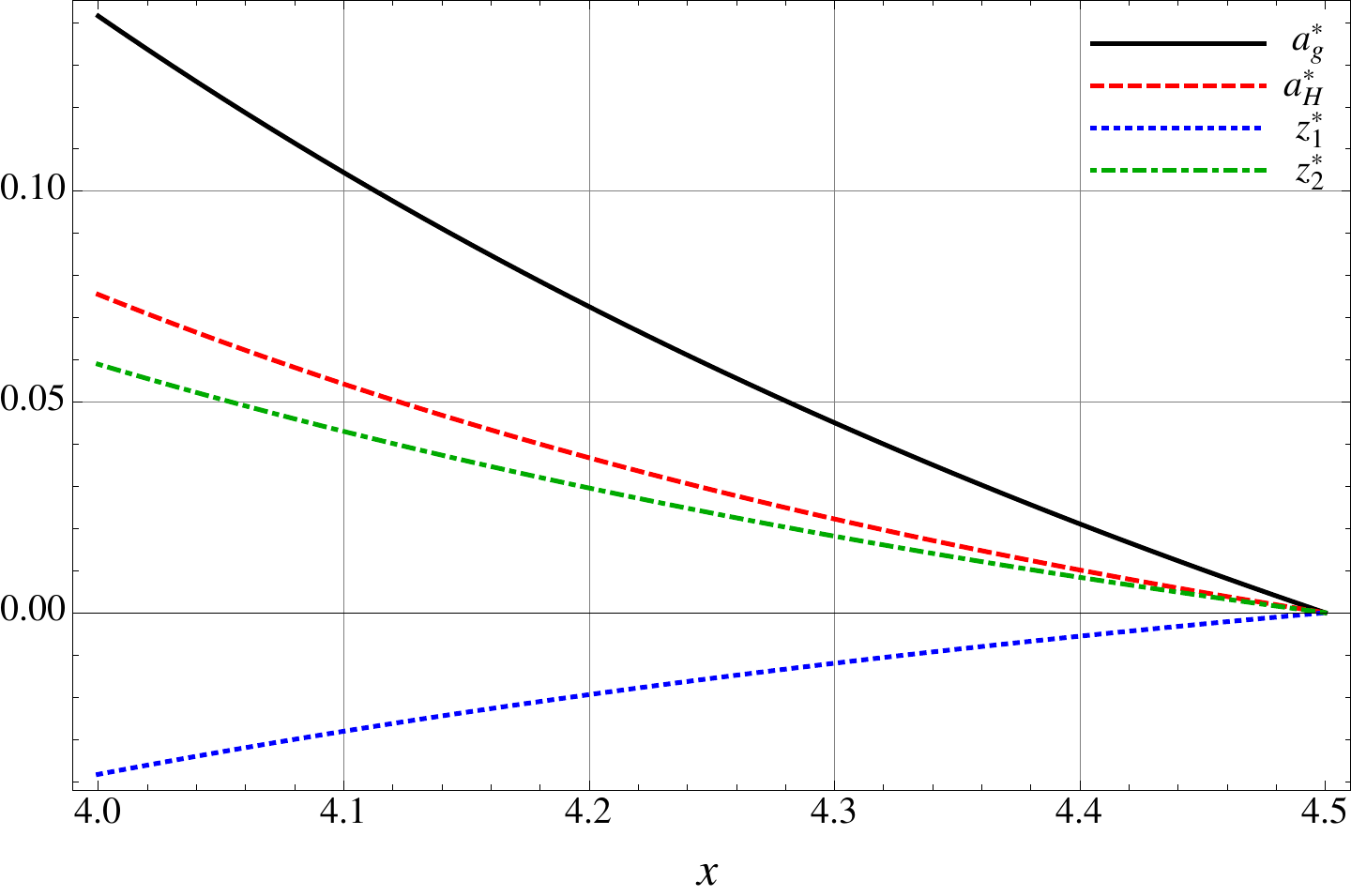}
\caption{The fixed points of the theory at $N_f=12$ as a function of $x$, the ratio between the number of flavors and number of colors.}
\label{amsfixed}
\end{center}
\end{figure}

To find the fixed points, we choose $N_f=12$\footnote{This is sufficiently large that the analysis is completely equivalent with the one for infinite $N_f$ and $N_{TC}$.} and set the beta functions equal to zero. We then plot the dependence of these fixed points on the ratio $x$ as depicted in Figure \ref{amsfixed}, where the perturbative nature of the fixed point is evident\footnote{There are three additional solutions. One with lower values of $z_1$ which is unstable and two with complex values of $z_1$ which are unphysical.}.

\begin{figure}[h]
\begin{center}
\includegraphics[width = 0.6\linewidth]{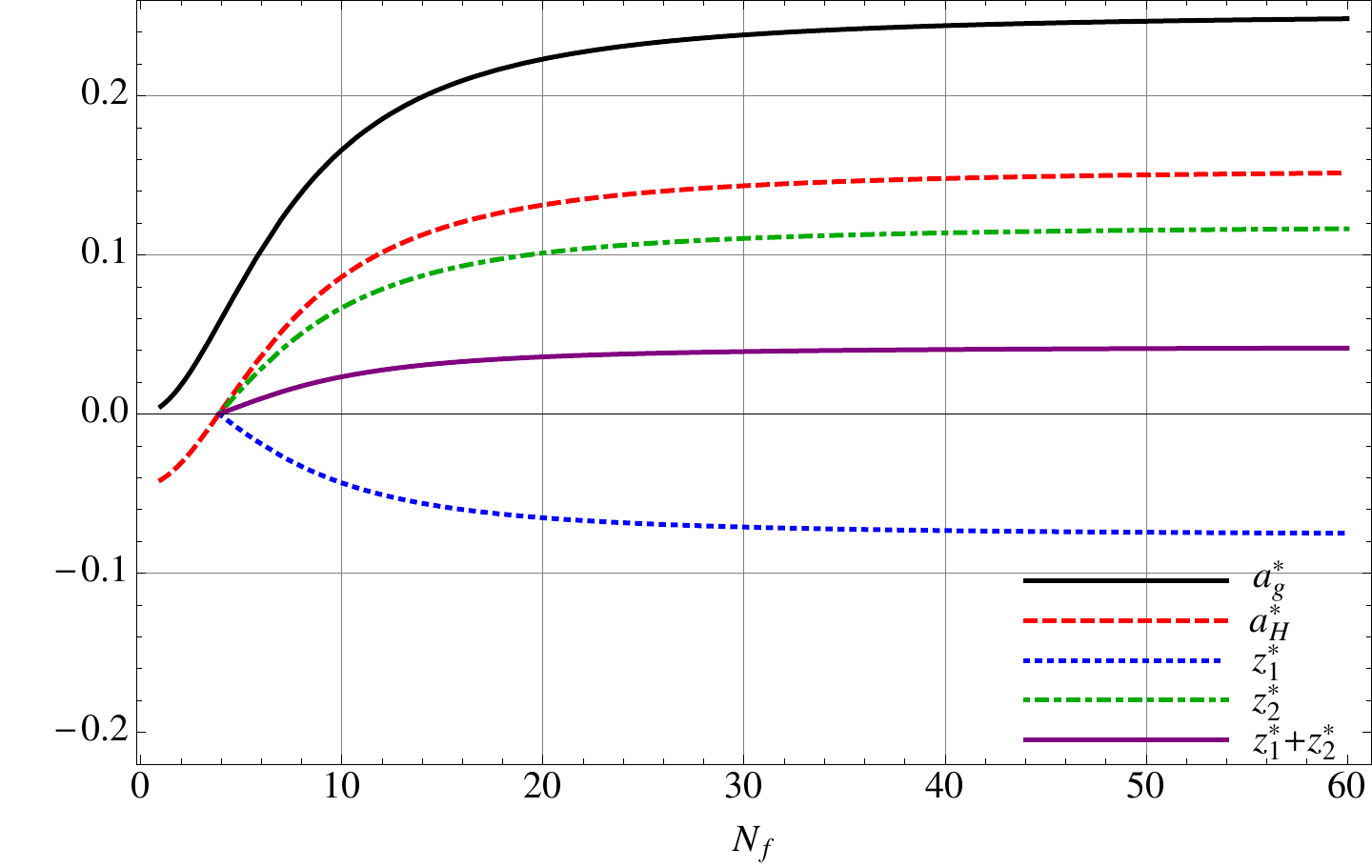}
\caption{The fixed points of the theory with $x=4$ as a function of $N_f$.}
\label{zzfixedxNf}
\end{center}
\end{figure}

To see that $N_f=12$ is a representative choice, we also plot the fixed point dependence on $N_f$ with $x$ held constant at 4 in Figure \ref{zzfixedxNf}.

Since the goal of this analysis is to investigate spontaneous breaking of chiral symmetry in the theory, we turn our attention to the dynamics of the scalar self-couplings, $z_1$ and $z_2$. As we have seen, the theory possesses fixed points in these couplings, and we note that the gauge and Yukawa beta functions \eqref{amsbetag} and \eqref{amsbetaH} do not depend on the quartic couplings. Therefore, we may set the gauge and Yukawa coupling constants to their fixed point values, while we let the quartic couplings evolve with the scale of the theory.

As the quartic couplings vary, the effective potential of the scalars changes, and it is possible to follow the RG flow into a region where the effective potential displays spontaneous symmetry breaking due to the one loop quantum corrections. Following the analysis of \cite{Gildener:1976ih}, the authors of \cite{Antipin:2011aa} calculated that the requirements for spontaneous symmetry breaking are
\begin{equation}
\label{amsreq1}
z_2 > 0 \hspace{1 cm} z_1 + z_2 < 0  \ . 
\end{equation}
This condition ensures that the potential is lower than that at the origin, such that a global minimum can appear. Furthermore imposing that the first derivative of the effective potential vanishes,
\begin{equation}
\label{amsreq2}
\beta_{z_1} + \beta_{z_2} + 4(z_1+z_2) = 0,
\end{equation}
forces the potential to have a nontrivial extremum which is a local minimum when 
\begin{equation}
\label{amsreq3}
 \sum_{\lambda_i\in\{a_g,a_H,z_1,z_2\}}\left[ \beta_{\lambda_i} \frac{\partial}{\partial \lambda_i} + 4\right](\beta_{z_1} +\beta_{z_2}) > 0.
\end{equation}

\begin{figure}[h!bt]
\begin{center}
\includegraphics[width = .6\linewidth]{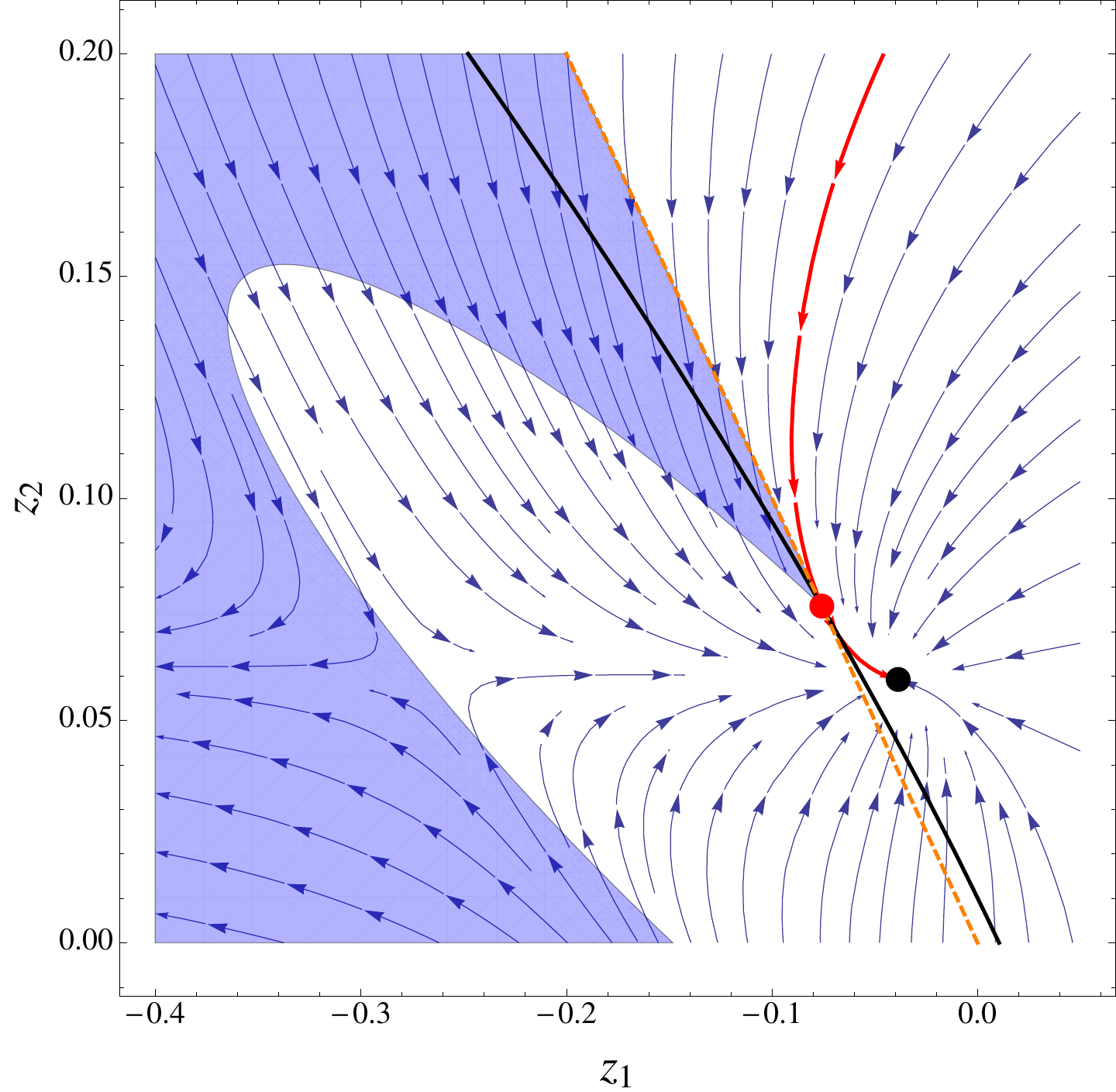}
\caption{The RG flows of the quartic couplings displaying the physical all directions stable fixed point for $x=4$ and $N_f=12$ with the conditions for spontaneous symmetry breaking displayed. The thick black line corresponds to the points where a nontrivial extremum appears. The blue region marks where this extremum is global minimum, i.e where the potential $V(\langle H \rangle) < 0$ and its second derivative $V''(\langle H \rangle) > 0$.  The red RG flow separates the two phases in the theory, one with symmetry breaking and one infrared conformal, and is called the separatrix. The dashed, orange line corresponds to the requirement that $z_1+z_2=0$.}
\label{amsflow}
\end{center}
\end{figure}

The region of coupling constant space where \eqref{amsreq1} and \eqref{amsreq3} are fulfilled is marked with blue in Figure \ref{amsflow}, while the requirement \eqref{amsreq2} gives the black stability line.
Interestingly, the theory displays two different phases, separated by the red separatrix; an infrared conformal phase, where the RG flows terminate at the fixed point, and one where the global $SU(N_f) \times SU(N_f)$ symmetry is spontaneously broken and a light scalar state arises as physical fluctuation along the diagonal of $H$. This scalar state is a pseudo-Goldstone boson associated to the breaking of conformal symmetry, and is known as \textit{dilaton} since its mass saturates the trace of the energy momentum tensor.

As was shown in \cite{Antipin:2011aa}, the mass of the dilaton arises at one loop level and is calculable from the one loop effective potential
\begin{align}
m_d^2 = \frac{2(4\pi)^2v^2}{N_f^2}\left[\frac{N_f^2-1}{N_f^2}4(z_2^0)^2-(a_H^0)^2x\right],
\label{amsdilaton}
\end{align}
where the zeroes indicate that the coupling constants are evaluated at a scale where $z_1^0 + z_2^0 = 0$. It is seen that the dilaton state may be tuned to arbitrarily small mass without turning off interactions. This is done by moving closer to the conformal phase bordered by the separatrix. The comparison with the masses of the other states of the theory was performed in \cite{Antipin:2011aa}. 

The dynamical features of this theory resemble those expected of a walking technicolor theory, a resemblance which is increased further when the analysis is taken to two loops in the beta functions \cite{Antipin:2012kc}, although a proper study of the actual order of the phase transition reveals an intriguing picture \cite{Sannino:2012wy,Antipin:2012sm}. 
 
\section{Electroweak embedding and standard model-like Yukawa interactions}\label{smImp}
The electroweak gauge sector is introduced by gauging the matter fields as follows
\begin{align}
\label{covds}
(D_{\mu}H)_{ij} &= \partial_{\mu}H_{ij} - ig'A_{\mu}^{r}S'^{r}_{ik}H_{kj} + ig''B_{\mu}H_{ik}S_{kj}'' \\
(D_{\mu}Q)_i &= \partial Q_i - igF^a_{ij}G_{\mu}^aQ_j -ig'A^{r}_{\mu}F'^{r}_{ij}Q_j\\
(D_{\mu}\tilde Q)_i &= \partial \tilde{Q}_i + igF^a_{ij}G_{\mu}^a\tilde{Q}_j - i g''B_{\mu}\tilde{Q}_jF''_{ji},
\end{align}
where $G_\mu^a$ is gluon field and $F^a_{ij}$ the associated field strength tensor.

The electroweak gauge transformations are governed by the coupling constants $g'$ and $g''$, the vector boson fields $A_{\mu}^r$ and $B_{\mu}$, and the generators $S$ and $F$. We gauge $N_f/2$ doublets with respect to the electroweak $SU(2)\times U(1)$ gauge symmetry and therefore define the following associated generators for the scalars 
\begin{eqnarray}
S'^r &=& \frac{\sigma^r}{2}\otimes\mathbb{1}_{N_f/2}
\\
S'' &=& \frac{\sigma^3}{2}\otimes\mathbb{1}_{N_f/2}\ ,
\end{eqnarray}
 and for the fermions \begin{eqnarray}
F'^r &=& S'^r \otimes \mathbb{1}_{N_{TC}}\\
F''&=& S'' \otimes\mathbb{1}_{N_{TC}}\ .
\end{eqnarray}

As has been shown elsewhere \cite{Antipin:2011aa}, the scalar potential is such that the ground state appears for a scalar vev of the form $\langle H \rangle = \tfrac{v}{\sqrt{2N_f}}\mathbb{1}$, the masses of the $W$ and $Z$ gauge bosons are given in the Table \ref{vbmass}. The physical masses of the $W$ and $Z$ bosons are recovered for $v=246 \text{ GeV}$. 
\begin{table}[hbt]
  \centering
   \begin{tabular}{| c | c | }
     \hline
     Vector boson & $m^2(v)$ \\ \hline 
     $W$ & $\frac{g'^2}{4}v^2$ \\ \hline
     $Z$ & $\frac{g'^2+g''^2}{4}v^2$ \\
     \hline
   \end{tabular}
  \caption{The tree-level masses of the weak vector bosons, assuming $\langle H \rangle = \tfrac{v}{\sqrt{2N_f}}\mathbb{1}$.}
  \label{vbmass}
\end{table}

In order to mimic the structure of the standard model, we furthermore wish to add a top quark-like interaction to the model. In order to do this, we add several quarks interacting with the scalars of the theory through a Yukawa coupling $y_q$ which we will later tune to reproduce the effect of a single top-like quark.

We expand the Lagrangian with a new Yukawa interaction
\begin{equation}
\Delta\mathcal{L}_q = -y_q\bar{q}_LHq_R
\end{equation}
where
\ea{
q_L = \left(\begin{array}{c}
q_L^1\\
q_L^2 \\
\vdots\\
q_L^{N_f}\end{array}\right)
;\qquad
q_R = \left(\begin{array}{c}
q_R^1\\
q_R^2 \\
\vdots\\
q_R^{N_f}\end{array}\right) \ .} 

These are ordered into $N_f/2$ generations of quark doublets of the form $(q^{\text{odd}}_L,q^{\text{even}}_L)$, each of which transforms in the fundamental of $SU(2)$ and with a standard model-like $U(1)$ hypercharge. Note that we have assigned the same Yukawa coupling to all of the new quark fields. To compensate for potential anomalies in the gauge sector, we should also add lepton-like fermions. We do not consider those here because their effects on the renormalization group analysis is subleading compared to the quark contribution, and because their effects on the potential can be safely neglected by assuming that their masses are small.

All of the newly introduced fields are singlets under the $SU(N_{TC})$ group, and transform in the fundamental representation of the QCD-like symmetry group.

We invoke the diagonal vev introduced earlier, and find
\begin{equation}
\Delta\mathcal{L}_q = -\frac{y_qv}{\sqrt{2N_f}}\bar{q}_L^i q^i_R,
\end{equation} 
such that we obtain a mass for the quarks given by 
\ea{
m_q= \frac{y_qv}{\sqrt{2N_f}}.
}
In order to simulate the effects of adding a single fermion with the mass of the top quark, $m_t=\frac{y_tv}{\sqrt{2}}$, and Yukawa coupling $y_t$, we require that the contributions to the one loop potentials are equal. At this level, every particle contributes to the effective potential with its mass to the fourth power \cite{Martin:2001vx}, this implies that 
\begin{equation}
N_fN_{QCD}m_q^4 = N_{QCD}m_t^4, 
\end{equation}
where $N_{QCD}=3$ is the number of colors in QCD, such that
\begin{equation*}
N_f\frac{y_q^4v^4}{4N_f^2}=\frac{y_t^4v^4}{4},
\end{equation*}
and we fix $y_q = N_f^{1/4}y_t$.

We have now completed introducing standard model-like effects into the theory, and can move on to finding the relevant corrections, with a focus on the RG dynamics.

\section{Standard Model-like impact on the phase structure and spectrum}\label{smCor}
We now wish to calculate the beta functions of the theory with standard model-like corrections, and to keep the notation consistent we introduce
\ea{
a_g' = \frac{g'^2}{(4\pi)^2}, \quad a_g'' = \frac{g''^2}{(4\pi)^2},\quad a_q = \frac{y_q^2N_{QCD}}{(4\pi)^2}.
} 
As mentioned above, we are considering fixed values of these new couplings, specifically we set 
\ea{
g'=0.6, \quad g''=0.3, \quad y_q = N_f^{1/4}\frac{\sqrt{2}m_t}{v}\approx N_f^{1/4}
}
with $m_t = 173$ GeV and $v=246$ GeV.

We add the standard model-like interactions with \emph{fixed} coupling strength given that we are interested in the physics near or at the electroweak scale, where we assume the chiral symmetry breaking dynamics controlled by the quasi fixed points (in absence of the standard model corrections) studied above to dominate. Of course, if we were to study  the model away from the electroweak scale one should consider the full set of beta functions which is not the focus of this work  \footnote{In addition we note that the electroweak-like gauge group $SU(2)\times U(1)$ does not respect the flavor symmetry of the scalar fields $H_{ij}$. The breaking of the symmetry that follows from this, means that $z_1$ and $z_2$ operators are no longer sufficient to parametrize the running of all interactions between these scalar fields. However, this effect is only brought about by terms in the beta functions proportional to $a_{g'}a_{g''}$, which is exceedingly small in the regions of parameter space we will consider  and can therefore be safely ignored.}. In fact, our main goal is to identify the impact of the standard model-like interactions on the fixed point analysis and dynamics at the electroweak scale. 

After adding the new sectors, we found the beta functions of the couplings and confirmed that they satisfy the Callan-Symanzik equation for the effective potential. They are
\begin{align}
\beta_{a_g} &= -2a_g^2\left[3-\frac{2x}{3}+\left(\frac{x^3}{N_f^2}+6-\frac{13x}{3}\right)a_g +x^2a_H -{\frac{x}{4}\left(3a_{g'}+a_{g''}\right) } \right] \\
\beta_{a_H} &= 2a_H\left[(1+x)a_H+ a_q-3\left(1-\frac{x^2}{N_f^2}\right)a_g- {\frac{3}{2}(3a_{g'} +a_{g''})} \right] \\
\beta_{z_1} &= 4\left[\left(1 + {\frac{4}{N_f^2}}\right)z_1^2+4z_1z_2+3z_2^2+z_1\left(a_H+a_q\right)-(3a_{g'}+a_{g''})z_1 + \frac{3N_f^2}{32}(a_{g'}^2+(a_{g'}+a_{g''})^2) \right] \\
\beta_{z_2} &= 4\left[{\frac{6}{N_f^2}z_1z_2}+2z_2^2+z_2\left(a_H+a_q\right)-\frac{x}{2}a_H^2  -\frac{N_f}{2N_{QCD}}a_q^2-(3a_{g'}+a_{g''})z_2\right],
\end{align}
where $N_f$ is the fixed, finite number of flavors.

\begin{figure}[bt]
\begin{center}
\includegraphics[width = 0.6\linewidth]{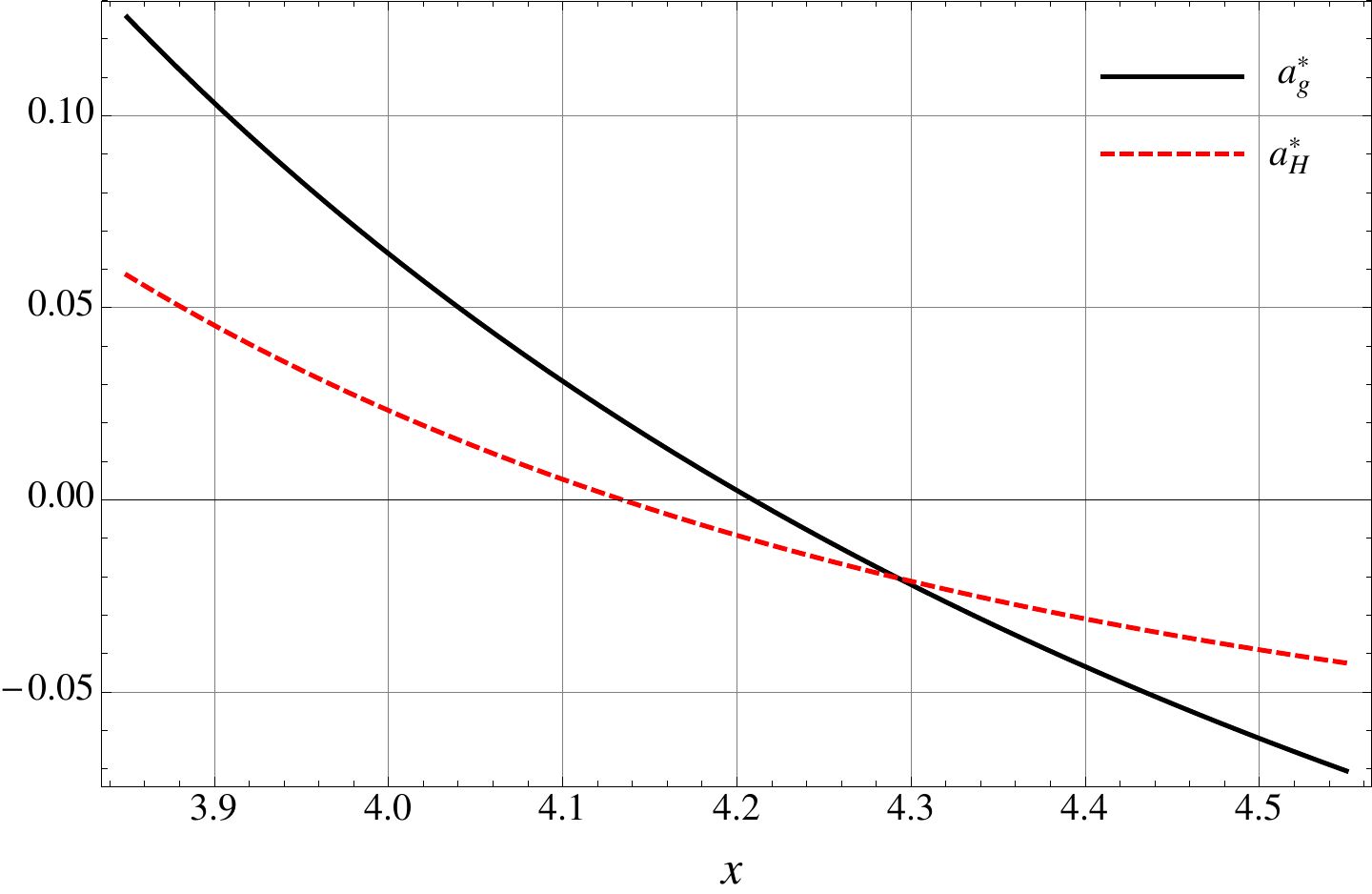}
\caption{Gauge and Yukawa fixed point values with $1/N_f^2$ and standard model-like corrections at $N_f = 12$ }
\label{gyfixed}
\end{center}
\end{figure}

The analysis in section \ref{amsSec} relied on the presence of fixed points in the gauge and Yukawa sectors. We now turn our attention to the dynamics of the corresponding coupling constants, which remain unaffected by the quartic couplings. They do, however, feel corrections from both the electroweak-like interactions and the finite $N_f$ corrections. The value of the coupling constants at the fixed point is shown in Figure \ref{gyfixed} for $N_f = 12$.
Evidently, the asymptotically free limit is affected by the standard model-like corrections. In addition we have to choose values of $x$ smaller than approximately 4.13 to ensure that the squared couplings are real at the fixed point.

\begin{figure}[bht]
\begin{center}
\includegraphics[width = 0.6\linewidth]{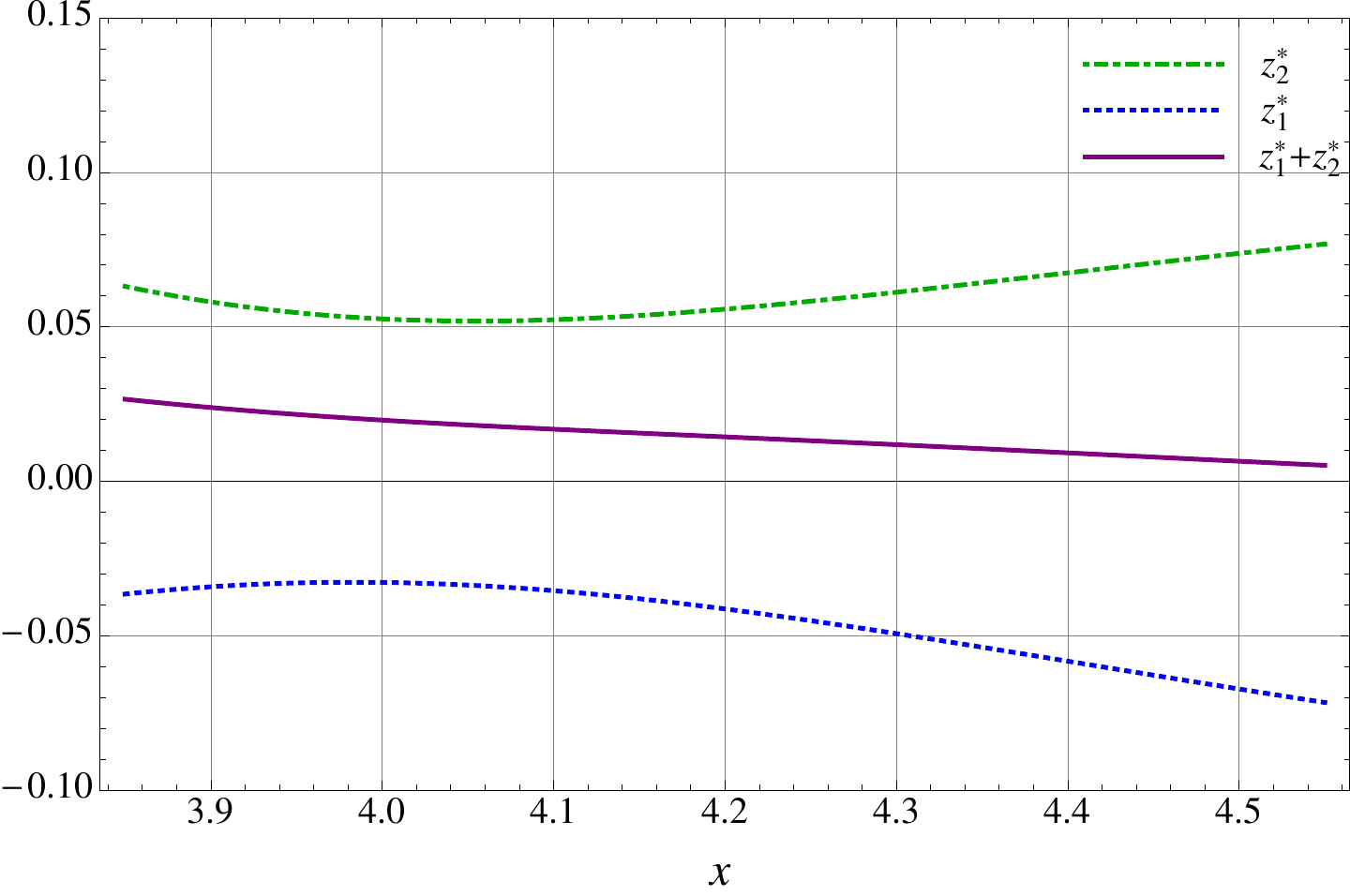}
\caption{The quartic IR stable fixed point values with the standard model-like corrections for $N_f = 12$ }
\label{zzfixed}
\end{center}
\end{figure}

With the established existence of a physical fixed point in the gauge-Yukawa sector, we tune these couplings to their fixed point values before studying the quartic beta functions. The effects on the fixed point in the quartic couplings are more severe, as depicted in Figure \ref{zzfixed}.
The main feature produced by the addition of standard model-like corrections is that there no longer is an $x$, where the fixed point values tend to zero. The smallest absolute values are obtained at $x\approx4$, where the sum is positive. Because, just as in Fig. \ref{amsflow}, the sum of the fixed point values is positive, the fixed point is on the right hand side of the line where $z_1+z_2=0$. This implies the existence of a conformal phase.
 
\begin{figure}[t]
\begin{center}
\includegraphics[width=.6\textwidth]{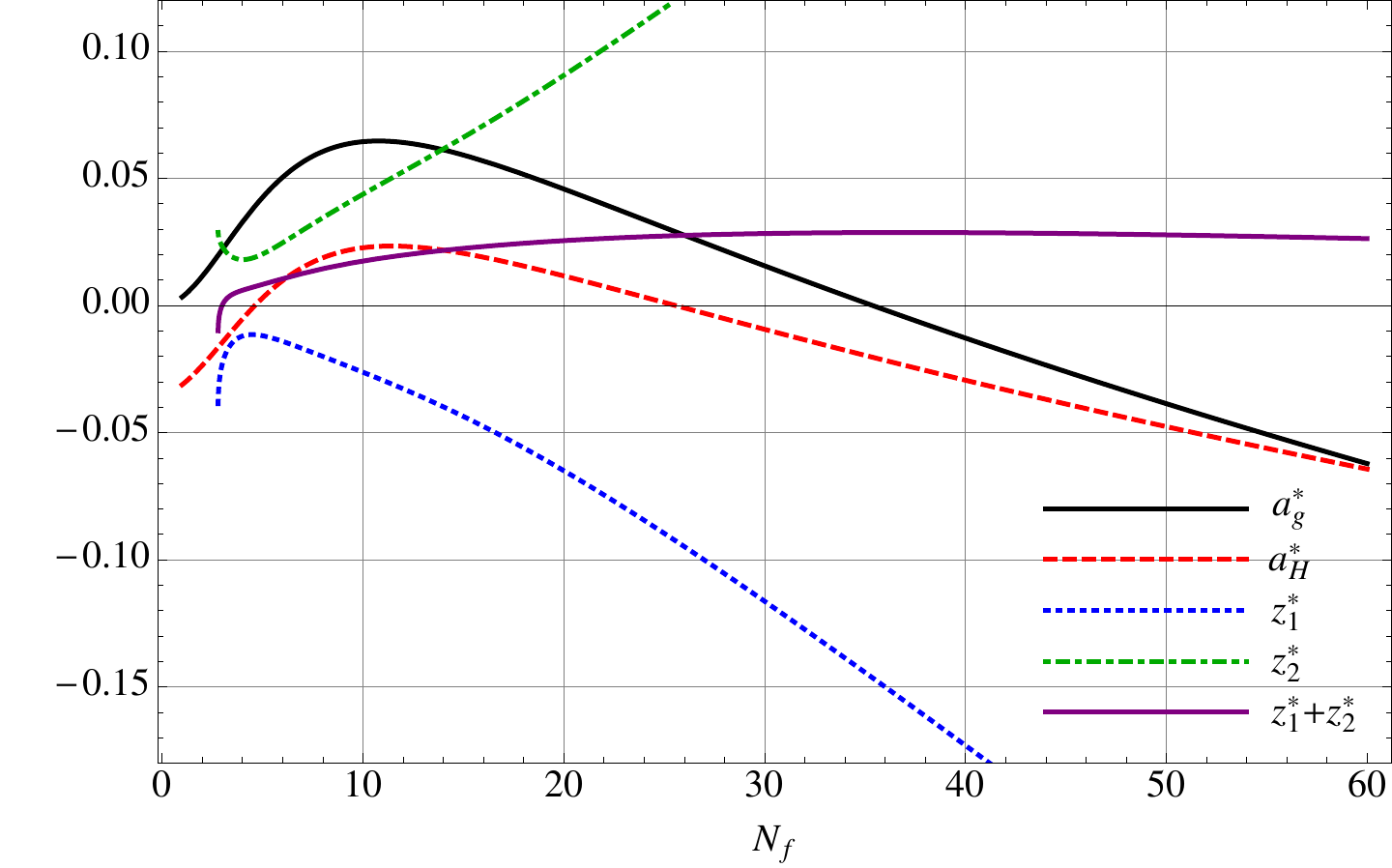}\label{zzfixedxSM}
\caption{The IR stable fixed point values with standard model-like corrections at $x = 4$ as function of $N_f$. }
\label{zzfixedN}
\end{center}
\end{figure}

As we have also introduced a specific $N_f$ dependence, we show the behavior of the fixed point values for different $N_f$ at a fixed $x$. In Figure \ref{zzfixedN} the $N_f$ dependence of the fixed points is shown for $x = 4$. For this value of $x$ we observe a wide range of $N_f$ values, from 5 to 25, where the fixed points exist. This range is controlled by $a_H^\ast$ which becomes negative, and therefore unphysical, outside it. 
Fixing $x$ evidently produces a conformal window between regions of negative Yukawa coupling. The high $N_f$ behavior is directly caused by the standard model-like terms of order $N_f$ and $N_f^2$ which turn up in the quartic beta functions, and drive them swiftly towards nonperturbative regions. For all allowed values of $N_f$, the sum of the values of the quartic couplings at the fixed points is positive, such that the theory has a conformal phase before considering the quantum corrections to the scalar potential.

\subsection{Phase structure}\label{PhaStr}

As we are mainly interested in the phase structure of the theory, and specifically whether or not spontaneous symmetry breaking occurs, we analyze the dynamics of the quartic couplings, assuming that the gauge and Yukawa couplings are set to their fixed point values.

\begin{figure}[h]
  \centering  
  \includegraphics[width=.47\textwidth]{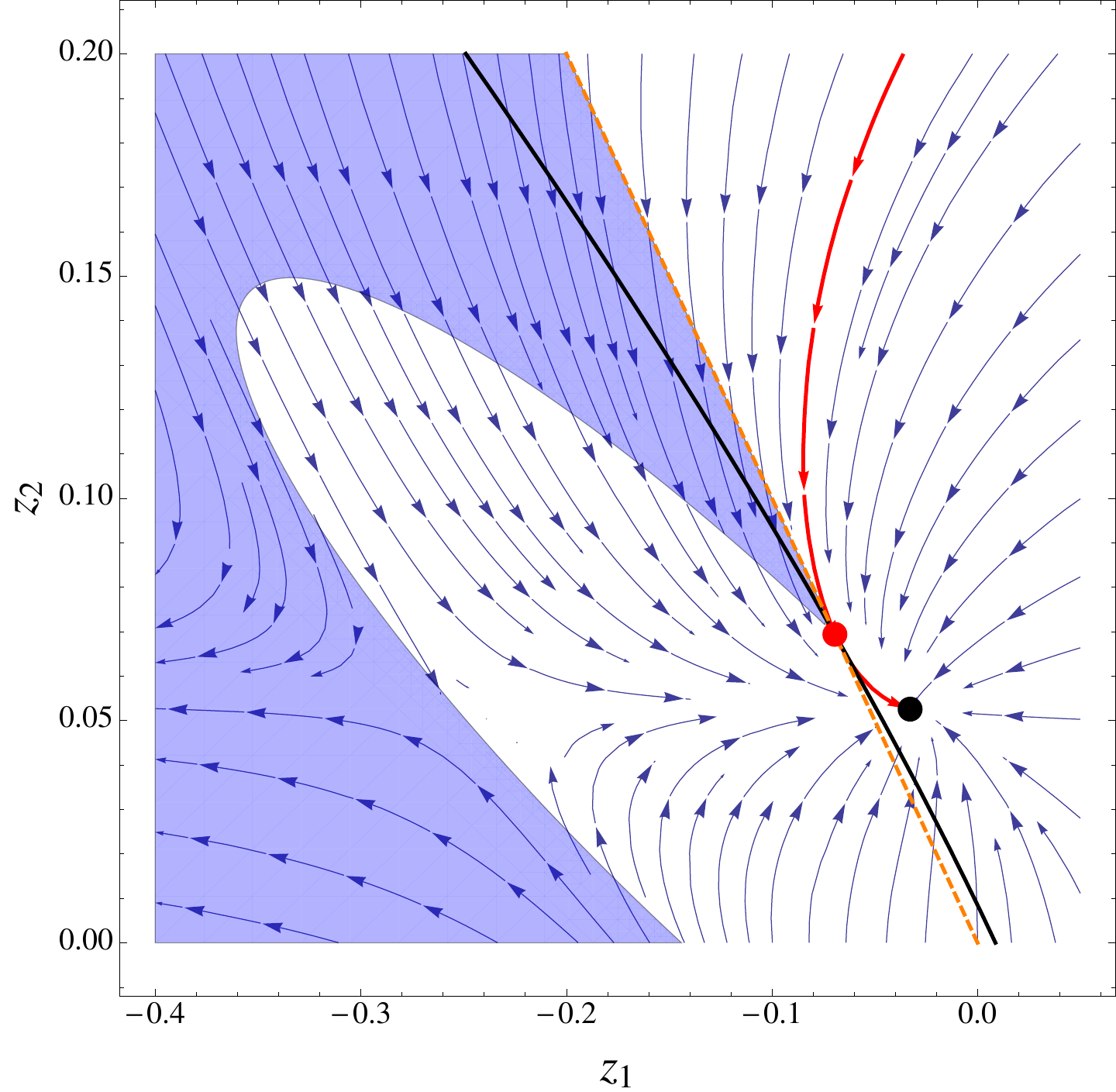}\
  \caption{The RG flows for the quartic couplings at $x= 4$ including the standard model-like corrections for $N_f = 12$. The basic features of the flows remain similar to the case in the absence of the standard model-like corrections, for this specific choice of parameters. We observe a conformal phase as well as one with spontaneous symmetry breaking.}
  \label{ewts01}
\end{figure}

In Figure \ref{ewts01} the RG flows of the quartic couplings are shown for $x=4$ with $N_f=12$ . 

For $N_f=12$, we recover a phase structure similar to the original theory, where a conformal phase and a phase of spontaneous symmetry breaking are separated by a specific RG flow, the separatrix. Thus, under this choice of parameter values, the qualitative features of the model are unchanged. Of course, in contrast to the analysis of Fig. \ref{amsflow}, we may only move along the flows for a short RG time so that our approximation of the fixed standard model couplings is valid. 

\subsection{Spectrum}\label{dilSpec}
With the addition of the massive vector bosons and the top quark the mass of the dilaton \eqref{amsdilaton} is modified and we find that
\begin{align}
m_{d}^2& = \frac{2(4\pi)^2v^2}{N_f^2}\left[\frac{N_f^2-1}{N_f^2}4(z_2^0)^2-(a_H^0)^2x\right] + \frac{1}{(4\pi)^2 v^2}\left[12 m_W^4 + 6 m_Z^4 -4N_{QCD}m_t^4\right].
\end{align}

\begin{figure}[h]
\begin{center}
\includegraphics[width = 0.5\linewidth]{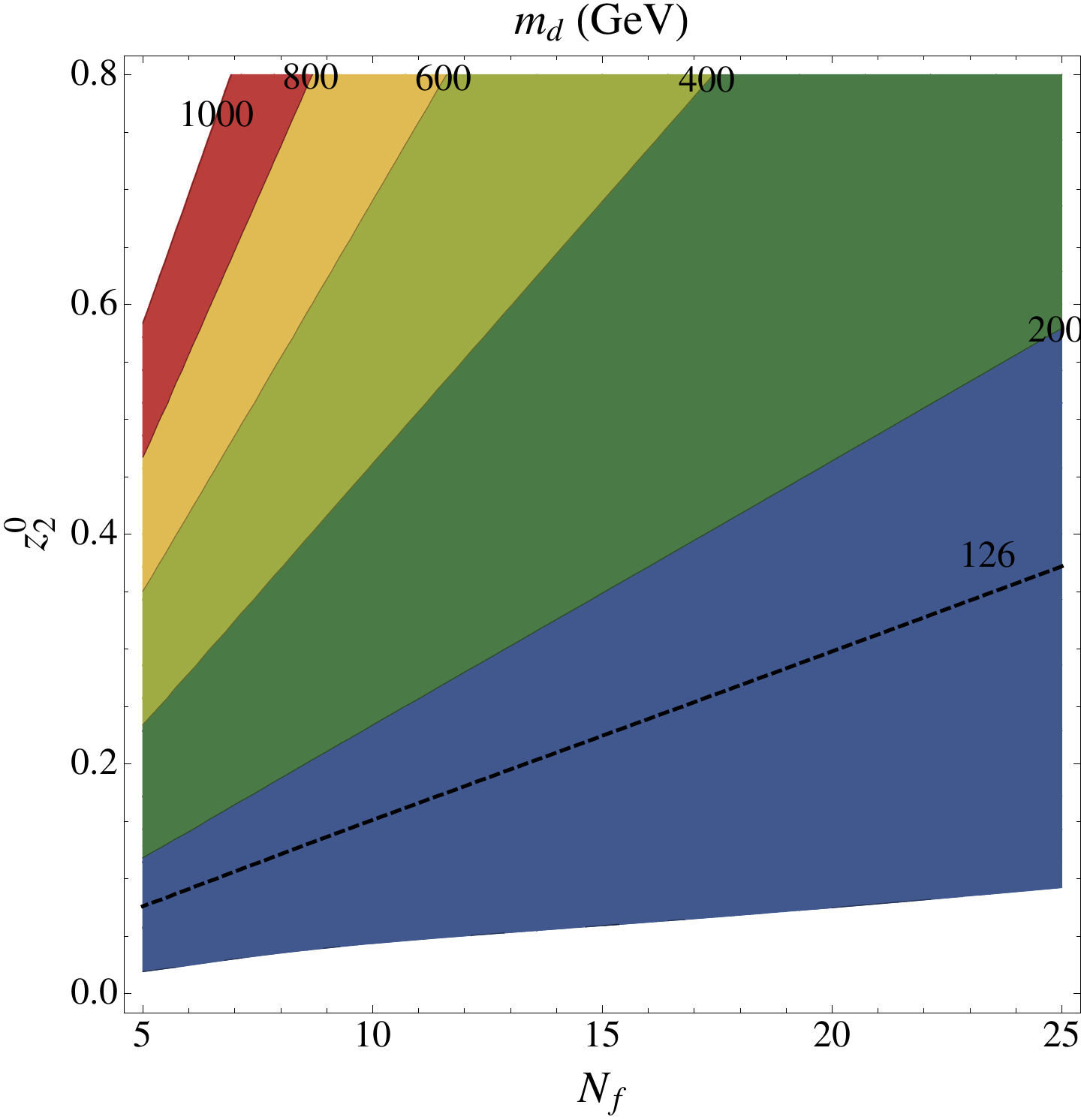} \hspace{0.7cm}
\caption{The mass of the dilaton as a function of the number of flavors and the $z_2$ coupling at $x=4$}
\label{dilmass}
\end{center}
\end{figure}

The mass of the dilaton is determined from $N_f$ and $x$ as well as the value of the $z_2^0$ coupling when spontaneous symmetry breaking occurs. In Figure \ref{dilmass} the behavior of the dilaton mass with respect to $N_f$ and $z_2^0$ is given for $x = 4$ with $a_H$ set to its fixed point value.
In this setup we see that we may reach $m_d \approx 126$ GeV when $z_2^0 = 0.18$ and $N_f =12$.  

It was shown in \cite{Antipin:2011aa} that in addition to the dilaton the theory possesses massless Goldstone bosons and a set of heavy scalars. The mass of these is
\ea{
m_S = 4\pi v \frac{\sqrt{2z_2^0}}{N_f} \ ,
}
and alas, we see that these values give a tree level mass of the heavy scalars of around 155 GeV. A heavier value of these masses can be obtained for a smaller value of $N_f$ still providing the expected Higgs mass.

\section{Conclusion}\label{conclusion}

We embeded the electroweak gauge group within the global flavor structure of a theory featuring perturbative near-conformal dynamics and a dilaton. Ordinary quark-like states have been added to mimic the top corrections. We have then examined the effects of the standard model-like interactions on the near-conformal dynamics of the model and shown that these corrections do modify the original phase diagram and the details of the dilatonic spectrum. In particular, we have shown that the corrected theory exhibits near-conformal behavior for a smaller range of flavors and colors. For this range of values our findings point to fact that near conformal dynamics could accommodate the observed  Higgs-like properties.

\acknowledgements
The CP$^3$-Origins centre is partially funded by the Danish National Research Foundation, grant number DNRF90.

\bibliographystyle{ieeetr}
\bibliography{dilabib}
\end{document}